\documentclass[reprint,amsmath,amssymb,aps,pra]{revtex4-1}

\usepackage{upgreek} % to have roman greek letter
\usepackage{epsfig, amsmath,amsfonts, amssymb,graphicx,amsthm,color}
\usepackage{mathtools}% For \MoveEqLeft
\usepackage{dcolumn}% Align table columns on decimal point
\usepackage{bm}% bold math
\usepackage{rotating}
\usepackage{units} % for \nicefrac{a}{b} <=> a/b
\usepackage{amssymb}
\usepackage{sidecap}
\usepackage[latin1]{inputenc}
\usepackage{graphicx}
\begin{document}
\title{Strong-field ionization of chiral molecules with bicircular laser fields : sub-barrier dynamics, interference, and vortices.}
\author{S. Beaulieu,$^{1}$ S. Larroque,$^{1}$ D. Descamps,$^{1}$ B. Fabre,$^{1}$ S. Petit,$^{1}$, R. Ta\"ieb,$^{2}$ B. Pons,$^{1}$ Y. Mairesse$^{1}$}
\affiliation{$^{1}$ Universit\'e de Bordeaux -- CNRS -- CEA, CELIA, UMR5107, Talence, France}
\affiliation{$^{2}$ Sorbonne Universit\'{e}, CNRS, Laboratoire de Chimie Physique -- Mati\`{e}re et Rayonnement, LCPMR, 75005 Paris, France}

\date{\today}
\begin{abstract}
Strong-field ionization by counter-rotating two-color laser fields produces quantum interference between photoelectrons emitted on the leading and trailing edges of the laser field oscillations. We show that in chiral molecules, this interference is asymmetric along the light propagation direction and strongly enhances the sensitivity of the attoclock scheme to molecular chirality. Calculations in a toy-model molecule with a short-range chiral potential show that this enhanced sensitivity already emerges at the exit of the tunnel. We investigate the possible sources of chiral sensitivity in the tunneling process, and find that the interference between electron vortices plays a crucial role in the chiral response. 
\end{abstract}

\maketitle
% figures made with /mnt/Documents/Data/Analyse/TOMO/BicircularBichromaticPaperFigures.m
% and /mnt/Documents/Data/Analyse/SFA/CompareCorotCounterrotSFA.m

\section{Introduction}
The dynamics of under-the-barrier particles in quantum tunneling has gained a lot of attention in the past few years \cite{ramos2020}. Strong-field light-matter interaction provides an interesting playground to investigate tunneling on the attosecond timescale \cite{kaushal2018,hofmann2019,kheifets2020}. When an atom or a molecule is submitted to a strong infrared laser field, the potential barrier evolves over a timescale of a few hundreds of attoseconds. The electrons can tunnel through the barrier when it is sufficiently low, and the dynamical evolution of the barrier dictates the properties of the emerging wavepacket -- momentum and phase \cite{lewenstein1994,ivanov2005}. If the direction of the laser field evolves in time (i.e. if the field is not linearly polarized), the 2D evolution of the barrier imprints vectorial properties on the tunneling electrons, determining the initial direction of their motion in the continuum. In that context, molecular chirality offers an interesting approach to reveal sub-barrier effects in strong-field ionization \cite{bloch2021}. 

When a chiral molecule is ionized by circularly polarized light, the ionic potential induces a forward-backward asymmetry in the ejected electron wavepacket along the light propagation direction, an effect called PhotoElectron Circular Dichroism (PECD) \cite{ritchie1975,powis2000,bowering2001,nahon2015,lux2012,lehmann2013,beaulieu2016}. In the single-photon ionization regime, PECD is imprinted during the scattering of the electron off the short-range chiral potential. In the strong-field ionization (SFI) regime, the electron is released though a sequential process : (i) tunneling through the potential barrier lowered by the laser field; (ii) acceleration by the laser field. The electron emerges from the tunnel a few atomic units away from the core, in a region where the ionic potential is mostly isotropic. However, PECD is still observed in strong-field ionization \cite{beaulieu2016,muller2020}, and can be imprinted through both steps of the SFI process. 

We have recently performed a joint experimental and theoretical investigation using the attoclock technique \cite{eckle2008} to decouple the influence of short and long-range effects in chiral photoionization \cite{bloch2021}. The attoclock consists in measuring the momentum distribution of photoelectrons produced by ionizing a target with a strong elliptically (or more complex polarization-shaped \cite{eicke2019}) laser field. The rotation of the laser field induces an angular streaking of the electron distribution, which maps the electron interaction with the ionic potential \cite{torlina2015,sainadh2019} and possibly its finite tunneling time \cite{hofmann2019}. In chiral molecules, experiments using elliptically polarized \cite{fehre2019b} or co-rotating bicircular laser fields \cite{bloch2021} established that the angular streaking produces different attoclock angles for electrons ejected forward and backward along the light propagation direction. This asymmetry was found to vanish in calculations when the long-range part of the ionic potential was screened \cite{bloch2021}, even though the number of electrons ejected forward and backward still differed. The attoclock thus seemed rather insensitive to short range effects in chiral photoionization.

In this article, we investigate the ionization of chiral molecules by bicircular bichromatic fields. In section \ref{section2}, we show that characteristic interference patterns appear in the photoelectron momentum distributions when the two components of the field are counter-rotating. Section \ref{section3} describes measurements of strong-field ionization of chiral molecules using co- and counter-rotating bicircular fields, which demonstrate the high sensitivity of the interference pattern to chirality. Simulations in a short-range chiral potential confirm the importance of interference in the chiral response, raising the question of the origin of the chiral sensitivity in the tunneling process. This question is addressed in Section \ref{section4}, using simulations within the Strong-Field Approximation (SFA) \cite{lewenstein1994} to investigate the sub-barrier dynamics in various field configurations -- circular, co-rotating and counter-rotating. We then model the chiral photoionization using different sources of chirosensitivity, and reveal the influence of interference of photoelectron vortices in counter-rotating fields. Last, in Section \ref{section5} we conclude our study by showing that changing the ratio between the two components of the bicircular fields can increase the contrast of the interference, resulting in very strong forward/backward asymmetries in the angular distributions close to destructive interference locations.

\section{Interference in counter-rotating bicircular photoionization}
\label{section2}
\begin{figure*}
\begin{center}
\includegraphics[width=0.8\textwidth]{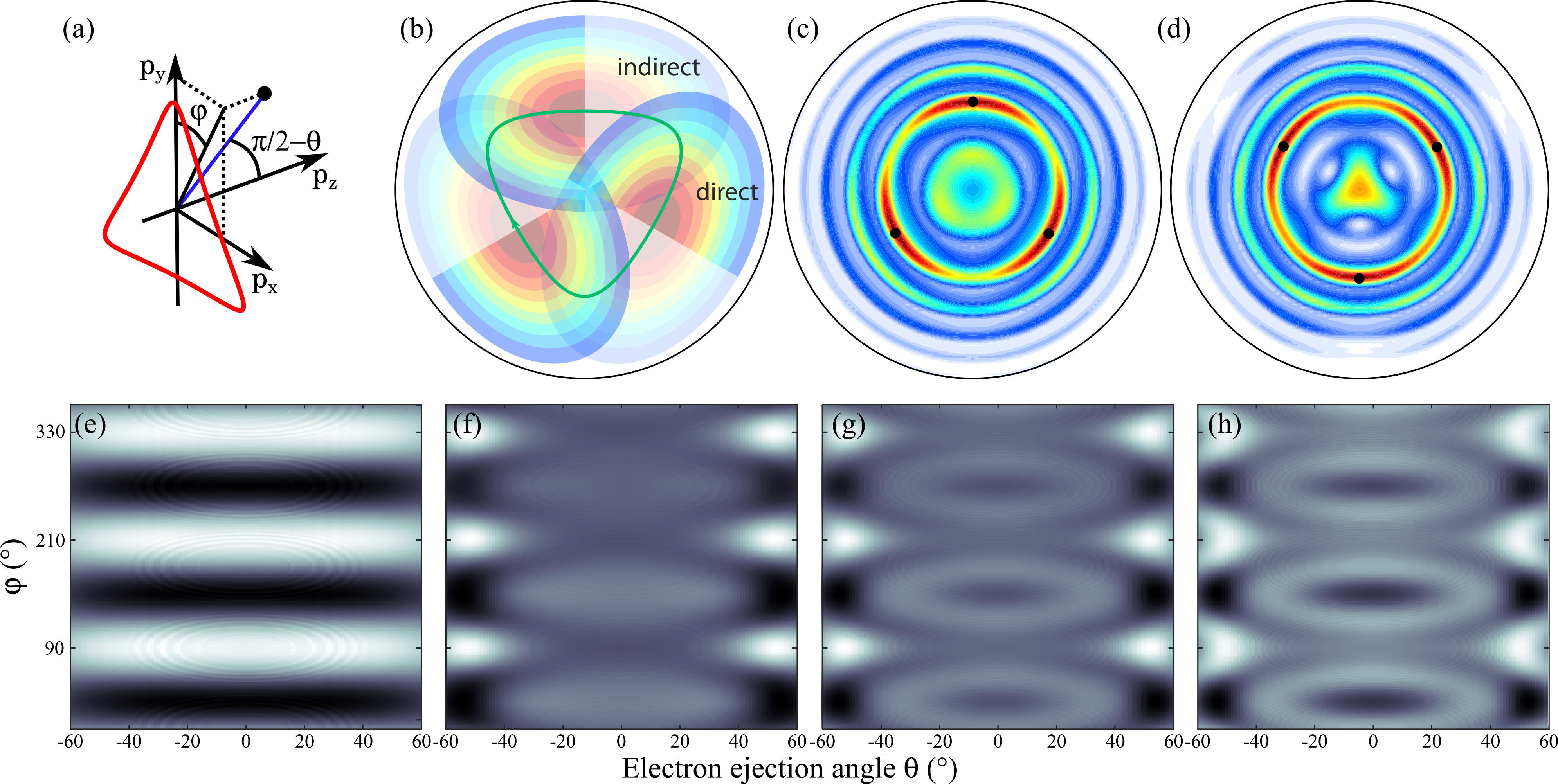}
	\caption{Principle of photoelectron interferometry in a counter-rotating bicircular field shown as red line in (a). (b) Modulus of the wavepackets released around each peak of the electric field, within the SFA, overlapped with the opposite of the vector potential (green). The direct and indirect parts of each wavepacket are depicted respectively in plain or transparent. (c-d) Photoelectron momentum distribution, projected in the laser polarization plane, obtained by incoherently (c) or coherently (d) summing the contributions of the three maxima of a four-cycle laser pulse. The dots mark the azimuthal maxima of the second ATI peak. (e-h) Azimuthal profile of the photoelectron momentum distribution as a function of the electron ejection angle $\theta$ with respect to the polarization plane, obtained incoherently (e) and coherently (f) for the second ATI peak, and coherently for ATI3 (g) and ATI4 (h). }
\label{FigWP}
\end{center}
\end{figure*}

Strong-field ionization generally releases two sets of electron trajectories. The electrons released before the maximum of the laser field move away from the core without turning back. By contrast the electrons released after the maximum of the field start their motion in one direction before reversing it, and can thus revisit the ionic core before being accelerated away by the laser field. These two families of electron trajectories are generally labeled "direct" and "indirect", respectively \cite{debohan2002}. In an elliptically polarized laser field, direct and indirect electrons are angularly separated by the rotation of the laser field, and end up with different final momenta. By contrast, in counter-rotating bicircular fields, the trajectories released by consecutive maxima of the laser field can overlap in momentum space, leading to a quantum interference process. To illustrate this, we consider a bicircular field consisting of a circularly polarized field at fundamental frequency $\omega=1.55$ eV (800 nm wavelength) and intensity $I_{\omega}=3\times10^{13}$ W/cm$^2$, and its counter-rotating second harmonic whose intensity $I_{2\omega}$ is defined by the ratio $r=I_{2\omega}/I_{\omega}=0.1$ (Fig. \ref{FigWP}(a)). Figure \ref{FigWP} shows the photoelectron momentum distribution obtained within the strong-field approximation  using the saddle-point method \cite{lewenstein1994}, for a target with 7.2 eV ionization potential. Since the influence of the ionic potential on the electron dynamics is neglected, the final momentum of the electron  $\mathbf{p}_{f}$ is dictated by the value of the vector potential $\mathbf{A}$ at the time of ionization $t_i$ and the initial momentum of the electron at the exit of the tunnel $\mathbf{p}_i$: $\mathbf{p}_{f}=-\mathbf{A}(t_i)+\mathbf{p}_{i}$. The counter-rotating laser field has a C$_3$ symmetry, maximizing three times per laser period. It consequently releases three wavepackets per period, whose momentum-resolved modulus in the laser polarization plane is shown in  Fig. \ref{FigWP}(b). The wavepackets follow the shape of $\mathbf{-A}(t)$ (depicted as a green line), and each wavepacket is divided in two parts, corresponding to direct and indirect electrons, released respectively before and after the maxima of the filed oscillations. The direct and indirect wavepackets from adjacent laser sub-cycles overlap in momentum space, and are thus expected to interfere. To determine the influence of this interference, we compare in Fig. \ref{FigWP}(c,d) the photoelectron angular distributions produced by a multicycle laser pulse, in which the contribution of the three third cycles are incoherently (b) or coherently (c) summed, i.e. obtained respectively by summing the modulus square of the three wavepackets or by calculating the modulus square of the sum of the complex wavepackets. The distributions are made of concentric rings, corresponding to the above-threshold ionization (ATI) peaks \cite{agostini1979}. The second ATI peak maximizes around azimuthal angle $\varphi_0=90^\circ [120^\circ]$ in the incoherent signal, and $\varphi_0=150^\circ [120^\circ]$ in the coherent signal (locations marked by dots  in Fig. \ref{FigWP}(c,d)). The interference between direct and indirect electrons thus completely reverses the azimuthal angular dependence of the photoelectron yield. The interference can be further studied by investigating the dependence of the electron azimuthal distribution as a function of the electron ejection angle $\theta$, defined in Fig. \ref{FigWP}(a). In the incoherent signal (Fig. \ref{FigWP}(e)), the position of the three lobes in the first ATI peak is independent of $\theta$. By contrast, in the coherent signal, a clear phase jump appears in the distribution around $\theta=45^\circ$. The interference also induces a splitting of the azimuthal profile of higher ATI peaks into 6 lobes.

This simple study demonstrates the importance of interference in strong field ionization by counter-rotating bicircular fields, even within a framework in which the influence of the ionic potential is neglected. The constructive or destructive nature of the interference is dictated by the phase between direct and indirect electrons. The different ionization dynamics of direct and indirect electrons is sufficient to induce a strong $\theta$-dependence of this phase, which transposes into a $60^\circ$ jump of the azimuthal distribution at a specific ejection angle $\theta$ for each ATI peak. Within the SFA, the lobes of the photoelectron azimuthal distribution are necessarily aligned the directions defined by the field, i.e. along $\varphi_0=90^\circ [120^\circ]$ or $\varphi_0=150^\circ [120^\circ]$. Beyond the SFA, the influence of the ionic potential on the departing electrons is expected to induce an angular shift $\varphi^{ion}(\theta)$ of the lobes, as observed in attoclock measurements \cite{eckle2008,torlina2015,sainadh2019,hofmann2019}, but should also affect the interference pattern because of scattering phase shifts.

\section{Chiral attoclock in co- and counter rotating fields}
\label{section3}
\subsection{Measurements}
In order to experimentally assess the sensitivity of the interferometric attoclock, we measured the 3D photoelectron angular distributions produced by photoionizing fenchone molecules with bicircular fields. We used the 1~kHz 25~fs Ti:Sa Aurore laser system at CELIA \cite{fedorov2020}. The bicircular field was generated in a Mach-Zehnder interferometer and focused in chiral molecules introduced in the interaction region of a Velocity Map Imaging Spectrometer as described in \cite{bloch2021}. The 3D photoelectron angular distribution of the electrons was reconstructed from 31 projections using a tomographic inversion method based on inverse Radon transform \cite{wollenhaupt2009,mancuso2015}. The chirosensitive part of the signal was extracted by subtracting the 3D photoelectron angular distribution obtained at a given helicity of the bicircular field to the mirror image of that obtained at opposite field helicity, as detailed in \cite{bloch2021}.

\begin{figure*}
\begin{center}
\includegraphics[width=\textwidth]{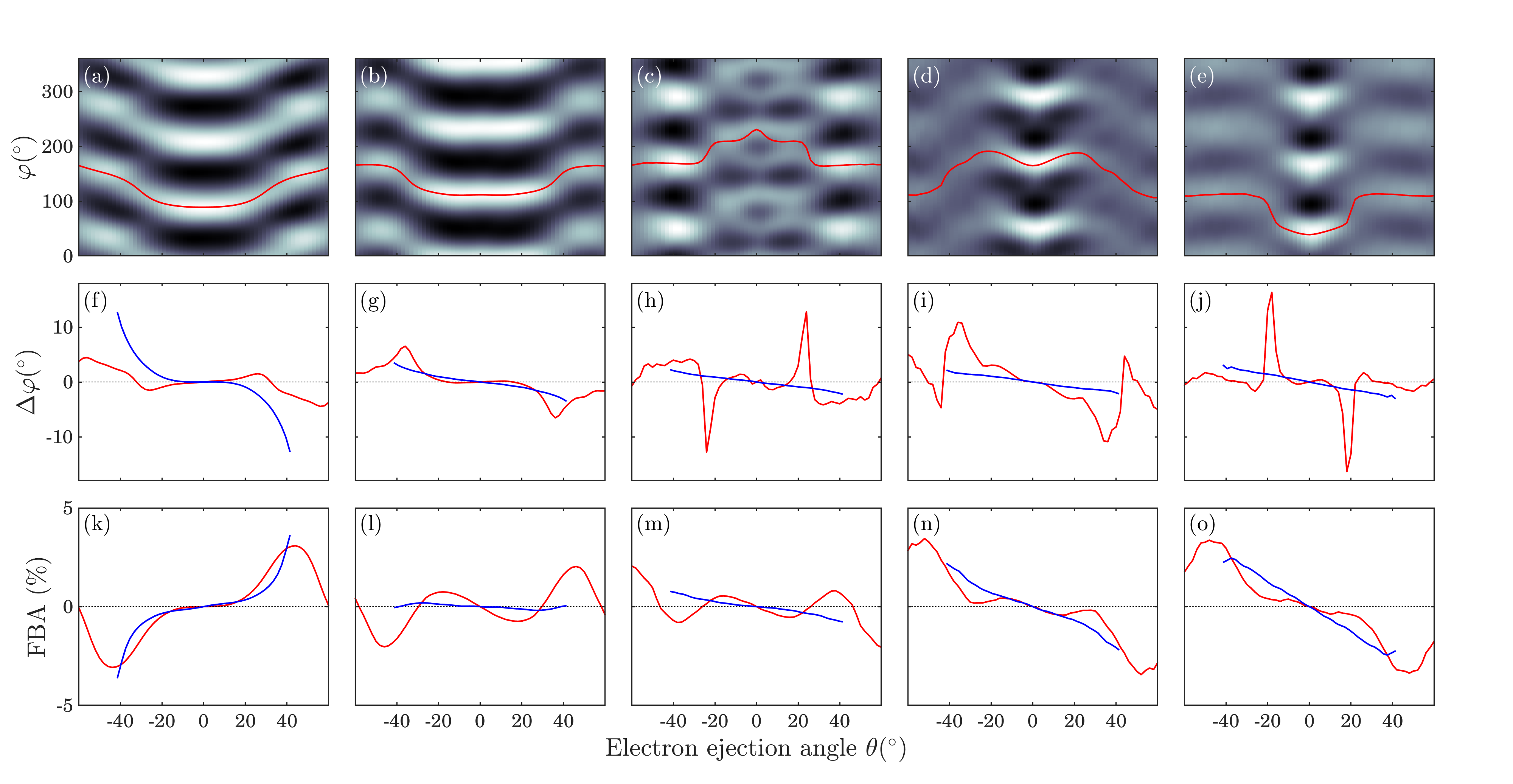}
	\caption{Interferometric attoclock measurements in (+)-fenchone using bicircular fields with $I_{\omega}\approx3\times 10^{13}$~W$\cdot$cm$^{-2}$, $r=I_{2\omega}/I_{\omega}=0.1$ and right circular polarization of the fundamental. (a-e) Photoelectron signal of the first five ATI peaks in the case of counter-rotating field configuration, as a function of the electron ejection angle $\theta$ and streaking angle $\varphi$. The signal is normalized at each $\theta$ by its $\varphi$-averaged value. The red lines represent the maximizing azimuthal angle $\varphi_0(\theta)$ extracted by Fourier analysis. (f-j) Chiral angular shift $\Delta\varphi_0(\theta)$ between electrons ejected forward and backward in counter-rotating (red) and co-rotating (blue) laser fields. (k-o) Forward/backward asymmetry in the photoelectron yield, using counter-rotating (red) or co-rotating (blue) fields. }
\label{FigResults}
\end{center}
\end{figure*}
Figure \ref{FigResults} shows the azimuthal profile of the first five ATI peaks produced in (+)-fenchone molecules using a counter-rotating bicircular field defined by $I_{\omega}\approx3\times10^{13}$ W.cm$^{-2}$ and $r=0.1$, as a function of the electron ejection angle $\theta$. At each $\theta$, the ATI yield is normalized by its $\varphi$-average value to enhance visibility. All ATI peaks show a clear threefold pattern along the azimuthal angle $\varphi$, corresponding to the ejection of three electron bunches per period of the field. The azimuthal profile of the first ATI peak is rather similar to the one obtained in SFA calculations, the lobes presenting a phase jump. The phase jump occurs at lower angles ($\theta\approx\pm30^\circ$) than in the SFA ($\theta\approx\pm45^\circ$) and its magnitude is larger than $60^\circ$. This reflects the sensitivity of the interferometric scheme to the influence of the ionic potential on the electron trajectories.  Higher ATI peaks show a more complex evolution as a function of the ejection angle. 

The chirosensitive part of the signal is obtained by comparing the angle $\varphi$ maximizing the azimuthal distribution (red line in Fig. \ref{FigWP}(a-e)) in the forward and backward directions, and calculating the chiral angular shift $\Delta\varphi_0(\theta)=\varphi_0(\theta)-\varphi_0(-\theta)$. All ATI peaks show a clear chiral angular shift, with a strong dependence on the electron ejection angle. For low ATI peaks, for which the evolution of the pattern is rather smooth, $\Delta\varphi_0$ is of the order of a few degrees, and switches sign across the phase jump in the interference. For higher ATI peaks, the chiral angular shift reaches high values (15$^\circ$ for ATI5), around the ejection angle at which a sharp phase jump is observed in the interference. This is due to the fact that the position of the phase jump slightly differs for electrons ejected in the forward and backward directions. For comparison, we plot the results of attoclock measurements with a \textit{co-rotating} bicircular field (blue lines in fig. 2 f-j), presented in \cite{bloch2021}. In this field configuration, the electron momentum distribution is sharper and the signal vanishes for $\theta>40^\circ$. For the first ATI peak, the chiral angular shift reaches values above 10$^\circ$ at $\theta\sim 30^\circ$, but it drops below 3$^\circ$ for higher ATI peaks. Within the investigated regime, the counter-rotating scheme thus possesses a superior sensitivity to chirality.

We can compare the attoclock measurements to a more conventional detection scheme which consists in determining the number of electrons ejected forward F ($\theta>0$) and backward B ($\theta<0$), integrated over $\varphi$, and calculating the Forward/Backward Asymmetry as FBA=2(F-B)/(F+B) (Fig. \ref{FigResults}(k-o)). The FBA signals obtained in co- and counter-rotating fields show the same order of magnitude and overall trends, but the counter-rotating case produces additional modulations of the FBA as a function of electron ejection angle, which we interpret as signatures of the interference process.  

\subsection{TDSE simulations using a short-range potential}
To shed light on the origin of the chiral sensitivity of SFI in bicircular fields, we performed quantum-mechanical calculations solving the Time-Dependent Schr\"odinger Equation (TSDE) using a toy-model chiral molecule. The details of the model can be found in \cite{bloch2021}. We employed the same laser field as in the experiments, defined by $I_{\omega}=3\times 10^{13}$~W$\cdot$cm$^{-2}$ and $r=0.1$, and carried out the same analysis as for the experimental data. In order to isolate the influence of short-range effects,  we damped the long-range chiral potential beyond a distance $r_0=3.5$ a.u. by multiplying it by an isotropic Yukawa cut-off  term $\exp^{-(r-r_0)}$ \cite{torlina2015,bloch2021}. 

\begin{figure*}
\begin{center}
\includegraphics[width=\textwidth]{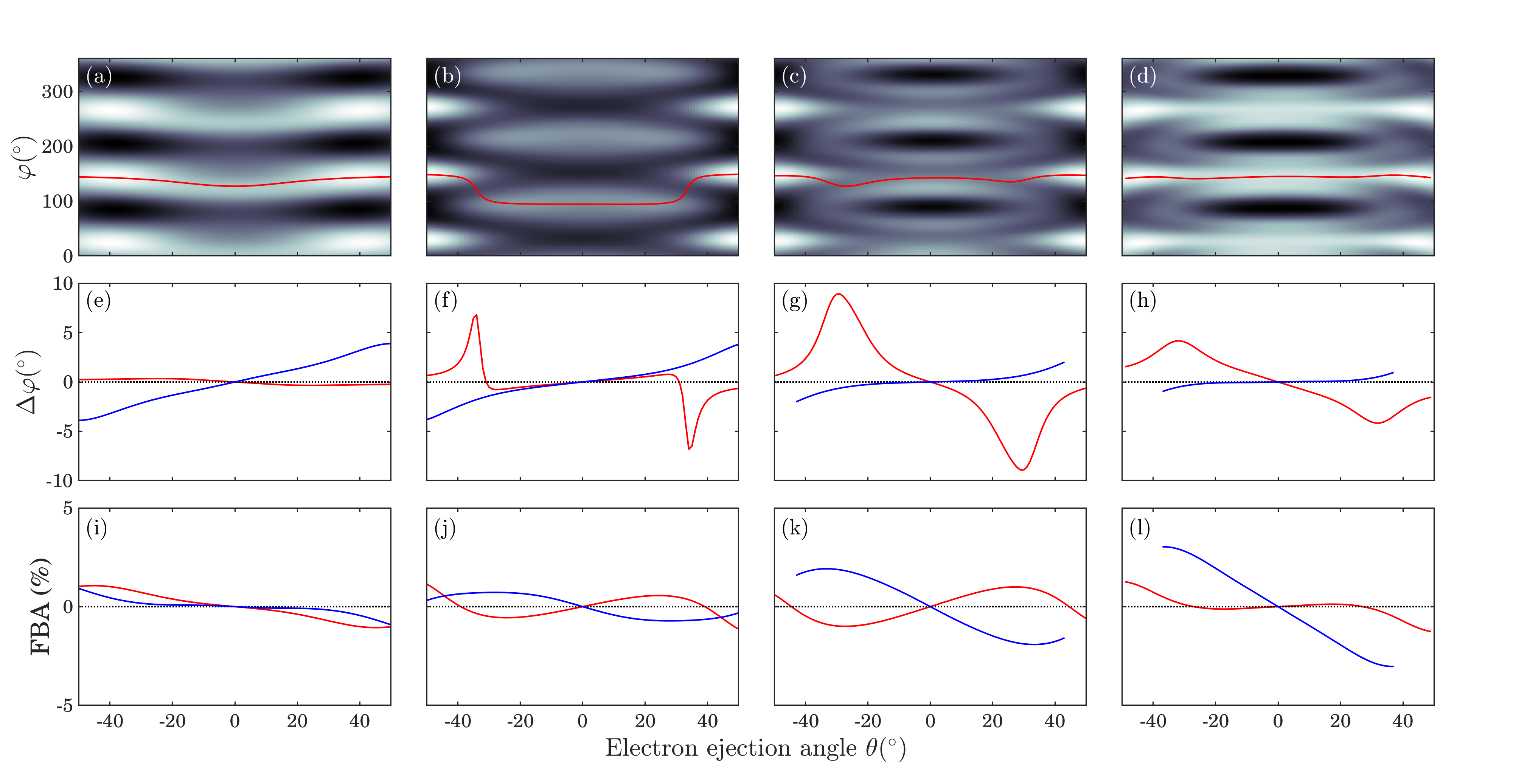}
	\caption{Interferometric attoclock calculations in toy-model chiral molecules with a screened long-range potential, subject to bicircular fields with $I=3\times 10^{13}$~W$\cdot$cm$^{-2}$, $r=I_{2\omega}/I_{\omega}=0.1$ and right circular polarization of the fundamental. (a-d) Photoelectron signal of the first four ATI peaks in the case of counter-rotating field configuration, as a function of the electron ejection angle $\theta$ and streaking angle $\varphi$. The signal is normalized at each $\theta$ by its $\varphi$-averaged value. The red lines represent the maximizing azimuthal angle $\varphi_0(\theta)$ extracted by Fourier analysis. (e-h) Chiral angular shift $\Delta\varphi_0(\theta)$ between electrons ejected forward and backward in counter-rotating (red) and co-rotating (blue) laser fields. (i-l) Forward/backward asymmetry in the photoelectron yield, using counter-rotating (red) or co-rotating (blue) fields.  }
\label{FigYukawa}
\end{center}
\end{figure*}

The angular distributions of the first four ATI peaks are shown in Fig. \ref{FigYukawa}. They resemble the results of SFA calculations (Fig. \ref{FigWP}), which reflects the efficiency of the potential screening by the above-described Yukawa cut-off term. Figure \ref{FigYukawa}(e-l) shows the analysis of the chiral attoclock, in co-rotating (blue) and counter-rotating (red) fields. When co-rotating fields are used, the chiral angular shift decreases with increasing electron kinetic energy ($\Delta\varphi_0<1^\circ$ for ATI3 and $\Delta\varphi_0<0.1^\circ$ for ATI4), as observed in \cite{bloch2021}. By contrast, in counter-rotating fields a clear angular shift is measured for all ATI peaks. The shift maximizes at the location of the phase jump in the interference pattern, as shown in Fig. \ref{FigYukawa}(a), reaching 9$^\circ$ for the third ATI peak. 

The forward/backward asymmetries are of the same order of magnitude in co- and counter-rotating field configurations, even if their signs are opposite for the intermediate ATI2-3 peaks. The asymmetry exhibits a sign switch in counter-rotating field configuration, at electron ejection angles where the chiral attoclock shows interference patterns. 

The persistence of a clear chiral angular shift in the SFI of molecules with a screened long-range potential is intriguing. In the absence of potential during the electron acceleration of the continuum, the azimuthal angle maximizing the electron distributions should be $\varphi_0=90^\circ$ or $150^\circ$, depending on the constructive or destructive nature of the interference between direct and indirect electrons. A phase shift between direct and indirect electrons emerging from the tunnel would change the position of the phase jump, but cannot rotate the distribution away from these $\varphi_0$ values within the SFA. On the other hand, a difference in amplitude between direct and indirect electrons would induce an effective rotation of the electron momentum distribution. To understand the origin of the interferometric chiral attoclock shift, we take a closer look at the SFA calculations by monitoring the electron trajectories.

\section{Strong-field electron dynamics in bicircular fields}
\label{section4}
\subsection{Electron trajectories in the complex plane}
\begin{figure*}
\begin{center}
\includegraphics[width=\textwidth]{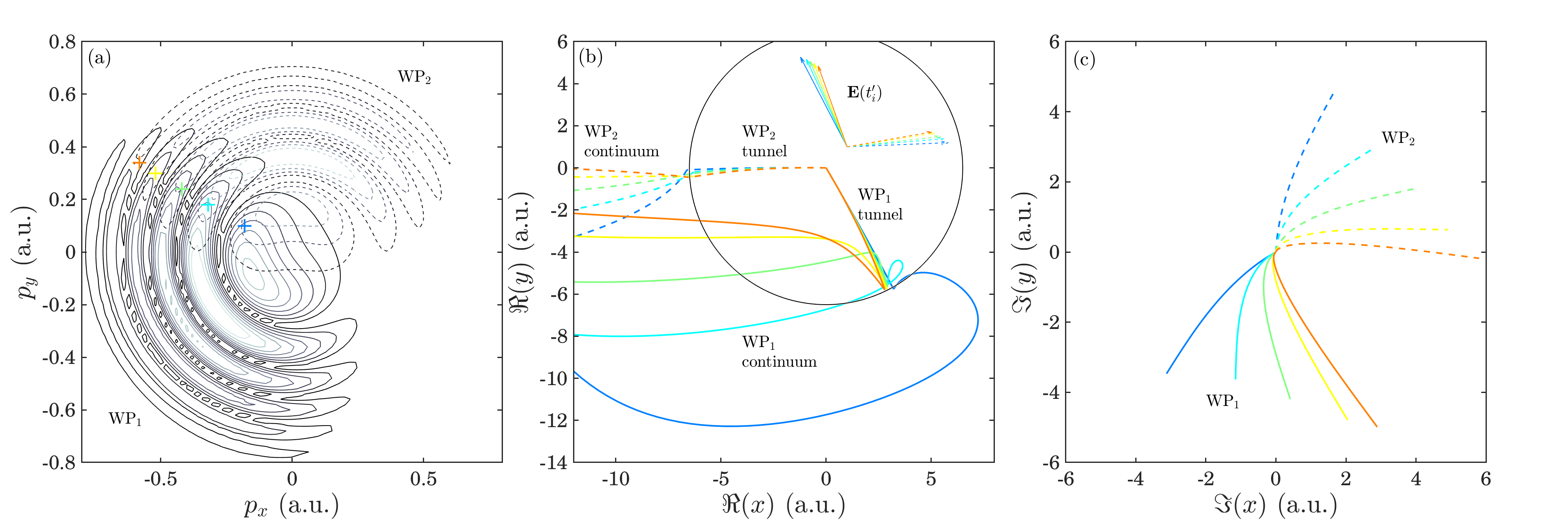}
	\caption{Electron trajectories in a counter-rotating bicircular field. (a) Final momentum distribution of the electron wavepacket produced by the first (continuous lines) and second (dashed lines) maxima of the laser field. (b) Real part of the electron trajectories producing different ATI peaks (from the first in blue to the fifth in orange), with a final momentum corresponding to the crosses in (a). The continuous lines are trajectories of the first wavepacket and are emitted on the trailing edge of the laser field oscillation. The dashed trajectories belong to the second wavepacket and are emitted on the leading edge of the field. The arrows represent the electric field at time $t_i'$ when the electron exits the tunnel. The black circle has a 6.5 a.u. radius and represents the approximate extent of the tunnel for the selected trajectories. (c) Imaginary part of the electron trajectories. The imaginary motion is confined to the sub-barrier region. }
\label{SFAtraj}
\end{center}
\end{figure*}

The SFA framework \cite{lewenstein1994a} provides a complete description of the electron motion, including the sub-barrier dynamics. The electron motion starts at time $t_i=t_i'+i t_i''$, where $t_i'$ and $t_i''$ are respectively the real and imaginary parts of the time. As time evolves along the imaginary axis from $t=t_i'+i t_i''$ to the real ionization time $t=t_i'$, the motion occurs under the potential barrier, in the classically-forbidden region, and is described in the complex plane \cite{smirnova2014}. The electron emerges from the tunnel with an initial velocity $\mathbf{v}(t_i')$ and is accelerated by the laser field to reach its final momentum $\mathbf{p_f}$. Figure \ref{SFAtraj}(a) shows the momentum distribution obtained by ionizing a target with 7.2 eV ionization potential by a counter-rotating bicircular field with $I_\omega=3\times10^{13}$ W.cm$^{-2}$  and $r=0.1$. We restricted the calculation to a single wavepacket emitted by the first (continuous lines, WP$_1$) or second (dashed lines, WP$_2$) third of the whole laser cycle, and repeated it over two consecutive cycles of the fundamental field to produce ATI peaks. The colored crosses mark the final momentum of electrons from the first five ATI peaks, which are produced with the same probability by the first or second wavepacket. The selected electrons belonging to the first wavepacket are emitted after the peak of the laser field, and are thus indirect, while the selected electrons belonging to the second wavepacket are direct. In order to monitor the differences between the electron ejection dynamics associated with both these wavepackets, we plot in Fig. \ref{SFAtraj}(b) the real part of their trajectories. The electrons start their motion in the sub-barrier region in the direction opposite to the laser field  $\mathbf{E}(t'_i)$ (represented as arrows), and exit the tunnel around 6.5 a.u. from the origin (radius of the black circle). The sub-barrier trajectories are slightly curved, with opposite handedness for direct and indirect electrons. The electron motion in the continuum is very different for the two families, in particular due to their different velocities at the exit of the tunnel. The velocity exit component $v_{\parallel}(t'_i)$, parallel to the laser electric field  $\mathbf{E}(t'_i)$, is positive for direct trajectories and negative for indirect ones. The transverse exit velocity $v_{\perp}$ changes sign between low and high order ATI peaks, but shows the same sign for direct and indirect electrons. In the imaginary plane (Fig. \ref{SFAtraj}(c)), where the electron motion is restricted to the sub-barrier region, the trajectories are strongly curved. Interestingly, the indirect imaginary trajectories are mirror images of the direct ones.  This is due to the fact that while direct and indirect electrons enter the barrier with the same imaginary angle, they are submitted to opposite forces because of a sign change of the imaginary component of the laser field, which results in opposite imaginary  torques.

\subsection{Characteristics of the complex electron motion}
\begin{figure*}
\begin{center}
\includegraphics[width=\textwidth]{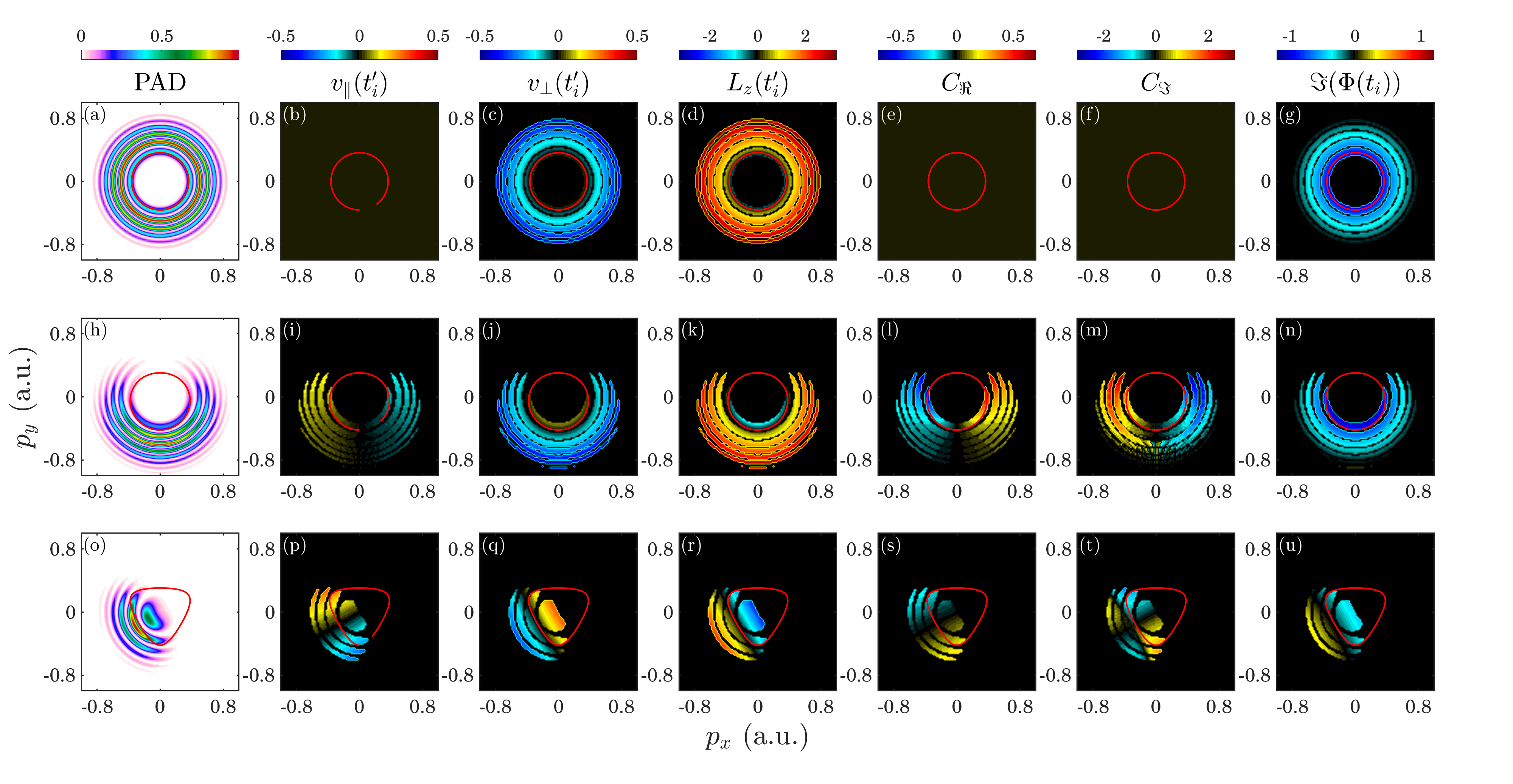}
	\caption{SFA calculations of photoionization of a target with 7.2 eV ionization potential by a circularly polarized 800 nm field at $3\times10^{13}$ W.cm$^{-2}$ (first row), a co-rotating bicircular field with $I_\omega=3\times10^{13}$ W.cm$^{-2}$  and $r=0.1$ (second row), and a counter-rotating field with the same parameters (third row). The first column corresponds to the photoelectron angular distribution in the polarization plane. 
The second and third columns represent the components of the exit tunnel velocities, parallel and perpendicular to the laser field, respectively. 
The fourth column depicts the electron angular momentum along the direction of light propagation as the electron enters the continuum. 
The curvatures of real and imaginary electron trajectories under the barrier are displayed in the fifth and sixth columns, respectively, while the last column shows the imaginary part of the entrance angle of the electron into the
barrier. }
\label{SFARecap}
\end{center}
\end{figure*}

In order to generalize the observations made on these few trajectories, we calculated the key parameters of the electron motion for photoionization by single color circularly polarized fundamental field, as well as co- and counter-rotating bichromatic fields. In the latter case we restricted the calculation to a single wavepacket emitted by one third of the whole laser cycle, to enhance visibility. The first column of Fig. \ref{SFARecap} (a,h,o) shows the photoelectron angular distributions in the polarization plane, and the next columns show the different parameters of the tunneling process that are discussed below. Note that a mask  is applied to represent the data only in the areas where the photoelectron yield in the polarization plane is above 10$\%$ of its maximum value. 

The first striking result that appears in Fig. \ref{SFARecap}(a,h,o) is that the electron distribution obtained in circular and co-rotating fields extends to higher momenta (i.e. is made of higher ATI peaks) than the one obtained in counter-rotating field. This is the result of the large velocity at the exit of the tunnel. This exit velocity is calculated by $\mathbf{v}(t_i')=\mathbf{p_f}+\mathbf{A}(t_i')$, and decomposed in components parallel $v_{\parallel}(t_i')$ (Fig. \ref{SFARecap}(b,i,p)) and perpendicular $v_{\perp}(t_i')$ (Fig. \ref{SFARecap}(d,k,r)) to the direction of the electric field $\mathbf{E}(t'_i)$. In the case of circular polarization (Fig. \ref{SFARecap}(b,c)), the exit velocity only consists of a perpendicular component which results from an outward centrifugal effect captured by uniform adiabatic asymptotics \cite{ohmi2015} and described by a Coriolis term in the rotating field frame \cite{dubois2024}. When a bicircular field is used (second and third rows of Fig. \ref{SFARecap}), a parallel velocity component appears at the exit of the tunnel. This velocity originates from the non-adiabaticity of the tunneling process due to the temporal variation of the magnitude of the electric field \cite{barth2011,ohmi2015,han2017,liu2019}. This has been observed in high-harmonic interferometry experiments, in which the velocity of the electrons emerging from the tunnel after the peak of the laser field has been measured \cite{pedatzur2015}. Our results show that the parallel exit velocity is negative for electrons ejected in an increasing 
laser field (direct electrons), and positive for electrons ejected in a decreasing laser field (indirect electrons). The indirect electrons thus start their motion in the continuum with a velocity opposed to the force acting on them. The parallel exit velocity reaches larger values in counter-rotating fields (up to 0.35 a.u.) compared to co-rotating fields (up to 0.1 a.u.), which is the sign of a prominent role of non-adiabatic effects in counter-rotating field configuration. By contrast, the perpendicular exit velocity is smaller in counter-rotating fields, as observed experimentally \cite{eckart2018a}, and changes sign across the classical $\mathbf{p_f}=-{\mathbf A}(t_i')$ curve. 

The perpendicular exit velocity is related to the angular momentum of the electron as it emerges in the continuum, ${\mathbf L}(t_i')={\mathbf r}(t_i')\times{\mathbf v}(t_i')$. The component of ${\mathbf L}(t_i')$ along $z$, $L_z(t_i')$, is shown in the fourth column of Fig. \ref{SFARecap}. $L_z(t_i')$, like $v_{\perp}(t_i')$, changes sign across the classical $\mathbf{p_f}=-{\mathbf A}(t_i')$ curve but is the same for direct and indirect electrons in counter-rotating field configuration. Therefore, this quantity does not enable to discriminate between electron trajectories at the exit of the tunnel in spite of the opposite handednesses of direct and indirect trajectories observed in Fig. \ref{SFAtraj}. We thus directly quantify 
these handednesses by calculating the normalized curvatures of the real and imaginary trajectories: $C_\Re=\int_{t_i'+i t_i''}^{t_i'}\left[\frac{\Re(\mathbf{r})}{\left|\Re(\mathbf{r})\right|}\times\frac{d\Re(\mathbf{r})}{dt} dt\right]$ 
and $C_\Im=\int_{t_i'+i t_i''}^{t_i'}\left[\frac{\Im(\mathbf{r})}{\left|\Im(\mathbf{r})\right|}\times\frac{d\Im(\mathbf{r})}{dt} dt\right]$, respectively. The results are shown in the fifth and sixth columns of Fig. \ref{SFARecap}. 
The curvature is more important in the imaginary plane, and switches sign between electrons ejected before and after the field maxima in bicircular field configurations. This is consistent with the trajectories displayed in Fig. \ref{SFAtraj}. 
An additional sign switch shows up in $C_\Im$ across the $\mathbf{p_f}=-{\mathbf A}(t_i')$ line in the case of counter-rotating fields, driven by the similar behavior of $v_{\perp}(t_i')$ shown in Fig. \ref{SFARecap}.  
Such a handedness switch in the imaginary electron trajectories produced by counter-rotating bicircular fields was previously observed by Ayuso and Smirnova \cite{ayuso2017}, who focused on recolliding (i.e. indirect) electron trajectories.
These authors related this switch to a change in the initial conditions of the electron motion -- a sign switch of the imaginary part of the electron entrance angle in the barrier $\Im(\Phi(t_i))=\Im(\text{atan}(v_x(t_i)/v_y(t_i)))$. 
We thus plot this imaginary angle in the last column of Fig. \ref{SFARecap}. The angle shows a sign change between low and high final momenta, matching the one observed in the handedness of the imaginary trajectories. However, $\Im(\Phi)$ has the same sign for direct and indirect electrons, whereas $C_\Im$ reverses.

\subsection{Modeling chiral photoionization}
The detailed analysis of the SFA dynamics sheds light on the possible sources of chirosensitivity of the tunneling process. In a previous work \cite{bloch2021}, we introduced a perturbed SFA model with a forward/backward asymmetric phase and amplitude modulation of the electrons emerging from the tunnel. Here we take a step further and investigate the dynamical origin of the process. Within a simple mechanistic picture of photoelectron circular dichroism, confirmed by classical calculations \cite{beaulieu2016}, the chiral potential converts the electron rotation motion in the polarization plane into translation along the propagation axis, in a similar manner as the thread of a nut converts the rotation of a bolt into translation \cite{powis2008}. Drawing from this, we can expect a chirosensitive momentum transfer $\delta p_z$ resulting from the rotation dynamics under the barrier. In this section, we estimate the effect of such a momentum transfer in co- and counter-rotating attoclock measurements. We introduce a momentum offset of the ionized electrons as $\delta p_z = \delta p_z^0 . \mathcal{M}$, where $\mathcal{M}$ is the source of chirosensitive momentum 
transfer. Among the sources explored in Fig. 5, we focus on the real ones, i.e. $L_z(t_i')$ and $C_\Re(t_i')$, since the imaginary ones, $\Im(\Phi(t_i))$ and $C_\Im(t_i')$, are rather linked to tunnel ionization probabilities as the standard imaginary part of the electron phase \cite{lewenstein1994,ayuso2017}. $\delta p_z^0$ is a proportionality factor such that each component of the 3D momentum distribution is shifted along $p_z$ according to the $\delta p_z^0$-scaled magnitude of $\mathcal{M}$ which depends on the electron trajectory. In counter-rotating fields, the three wavepackets emitted by the three thirds of the laser period are shifted independently. The wavepacket momentum distributions are interpolated on the same momentum grid, and the total distribution is obtained by coherently summing the interpolated contributions. 

\begin{figure*}
\begin{center}
\includegraphics[width=\textwidth]{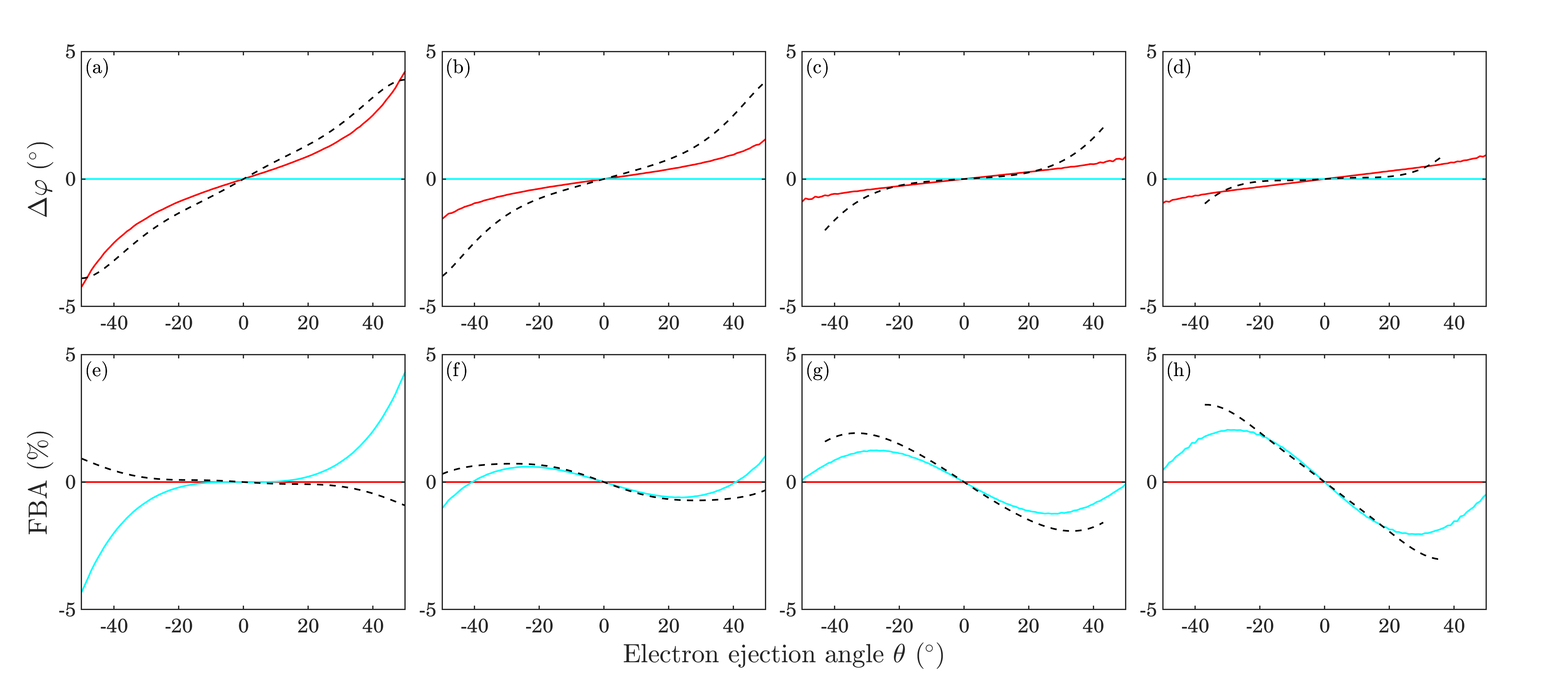}
	\caption{SFA-based simulations in a co-rotating bicircular field with $I_\omega=3\times10^{13}$ W.cm$^{-2}$ and $r=0.1$. Chiral attoclock shift (a-d) and forward/backward asymmetry in the electron yield (e-h) using a chirosensitive momentum shift due to the real trajectory curvature under the barrier, $\mathcal{M}=C_\Re$ (red), or to the angular momentum at the exit of the tunnel, $\mathcal{M}=L_z(t_i')$ (blue). The black dashed line is the result of the TDSE calculation in screened chiral molecules. }
\label{FigSimulCorot}
\end{center}
\end{figure*}

Figure \ref{FigSimulCorot} shows the results of the simulations obtained in a co-rotating field, in which no interference effect is at play. The main conclusions from the TDSE calculations, reproduced as dashed lines in Fig. \ref{FigSimulCorot}, were that the chiral attoclock shift almost vanishes for high enough energy electrons, i.e. when the potential screening is efficient, but that a forward/backward asymmetry in the electron yield persist, and even increases with electron energy. Let us examine the ability of the SFA-based calculations to reproduce these trends. If a chirosensitive momentum transfer is induced by the curvature of the sub-barrier trajectories, i.e. $\mathcal{M}=C_\Re$, then the opposite curvatures of direct and indirect electrons produce opposite momentum shifts. This induces a forward/backward asymmetry in the attoclock angle, visible in Fig. \ref{FigSimulCorot}. For these calculations, the value of $\delta p_z^0$ is such that the maximum momentum transfer experienced by the electrons is $\delta p_z^{max}\approx 10^{-3}$ a.u.. The chiral attoclock shift reaches $-4^\circ$ for electrons of the first ATI peak ejected around $-50^\circ$, and decreases with increasing electron kinetic energy. This overall trend is in agreement with the results of the TDSE calculations in the chiral molecules with a screened long-range potential, shown as dashed line in Fig. \ref{FigSimulCorot}(a-d). However, the opposite curvatures of direct and indirect electrons result in a compensation effect in the forward/backward asymmetry in the number of ejected electrons: the FBA is zero for all ATI peaks (Fig. \ref{FigSimulCorot} (e-h)). By contrast, assuming that the chirosensitive momentum transfer is proportional to the electron angular momentum at the exit of the tunnel, i.e. $\mathcal{M}=L_z(t_i')$, produces the same modulation for direct and indirect electrons, and thus no attoclock shift (Fig. \ref{FigSimulCorot} (a-d)) but a forward/backward asymmetry in the electron yield (Fig. \ref{FigSimulCorot} (e-h)). The FBA reaches $-4\%$ at $-50^\circ$ ejection angle for the first ATI peak, and switches sign for higher ATI peaks. For ATI4, it maximizes at $-28^\circ$ with value of $2\%$. For the two highest ATI peaks, the results are in qualitative agreement with the results of the TDSE calculations in screened chiral molecules. Assuming that a chirosensitive momentum shift is imprinted because of the sub-barrier electron rotation motion, proportional to the angular momentum at the exit of the tunnel, thus seems to provide a reasonable description of chiral tunneling. The situation would be qualitatively similar if we assumed that the imaginary entrance angle in the tunnel was the source of chirosensitivity, as proposed by \cite{ordonez2018} and recently used by Hofmann \textit{et al.} to describe chiral photoionization in bicircular fields \cite{hofmann2024}. 

\begin{figure*}
\begin{center}
\includegraphics[width=\textwidth]{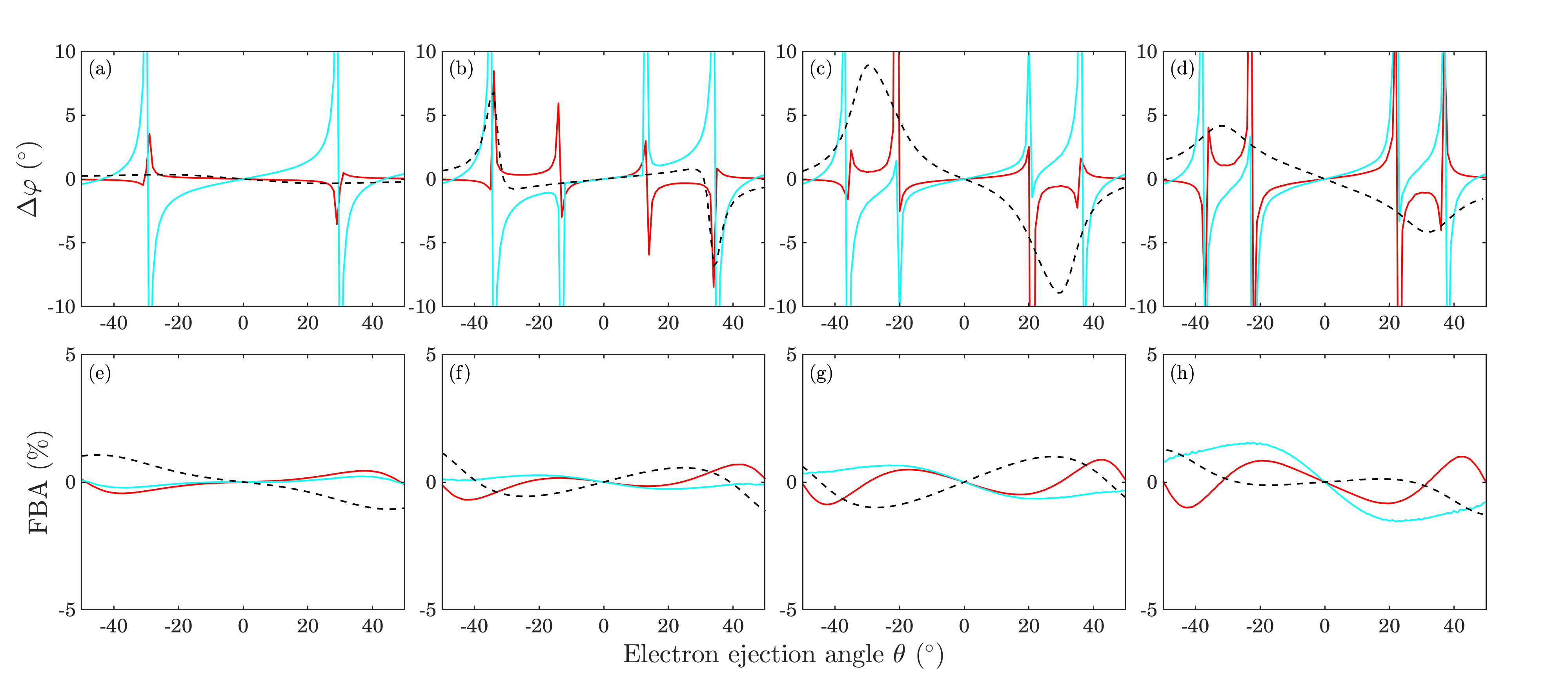}
	\caption{SFA-based simulations in counter-rotating bicircular fields. Chiral attoclock shift (a-d) and forward/backward asymmetry in the electron yield (e-h) using a chirosensitive momentum shift $\mathcal{M}=C_\Re$ (red) or $\mathcal{M}=L_z(t_i')$ (blue). The black dashed line is the result of the TDSE calculation in chiral molecules with a screened long-range potential. }
\label{FigSimulCounter}
\end{center}
\end{figure*}

We repeated the calculations using counter-rotating fields and keeping the same values of $\delta p_z^0$. The results, shown in Fig. \ref{FigSimulCounter}, present a striking feature: while in the co-rotating case, the modulation produced either a chiral attoclock shift but no FBA ($\mathcal{M}=C_\Re$ hypothesis), or a FBA but no chiral attoclock shift ($\mathcal{M}=L_z(t_i')$ hypothesis), we observe a chiral attoclock shift \textit{and} a forward/backward asymmetry in the photoelectron yield in both cases. The signals produced by the two hypothesis are indeed qualitatively similar. The attoclock shift shows a quasi-linear slope at small electron ejection angles. For ATI 2, the slope has the same sign as the one obtained in the TDSE calculations, but the signs are opposite for higher ATI peaks. Jumps are observed at well-defined values of $\theta$, where $\Delta \varphi$ exceeds 10$^\circ$. The jumps are sharper than the ones observed in the TDSE calculations, and two jumps are present for ATI 2-4 in the SFA-based model, while a single, broader one is observed in TDSE. We believe that this discrepancy is due to orientation-averaging effects. Our SFA-based model assumes that the influence of the chirality in tunneling results in a momentum transfer $\delta p_z^0$. In practice it is likely that different molecular orientations produce different values of $\delta p_z^0$, in a similar manner as the chiral response in photoionization depends on molecular orientation \cite{tia2017,fehre2019a}, which leads to a broadening of the response. Regarding the FBA, the order of magnitude of the SFA-based calculation matches the values observed in TDSE, but the signs and detailed evolutions are not well reproduced. In a previous study, we have shown that the phase of the wavepackets emerging from the tunnel was influenced by the chirality of the molecular potential \cite{bloch2021}. This forward/backward phase asymmetry of the wavepackets is not taken into account in the present model, and is likely to explain the discrepancy between the SFA-based calculations and the TDSE in counter-rotating fields, where interference effects play an essential role. 

\begin{figure*}
\begin{center}
\includegraphics[width=\textwidth]{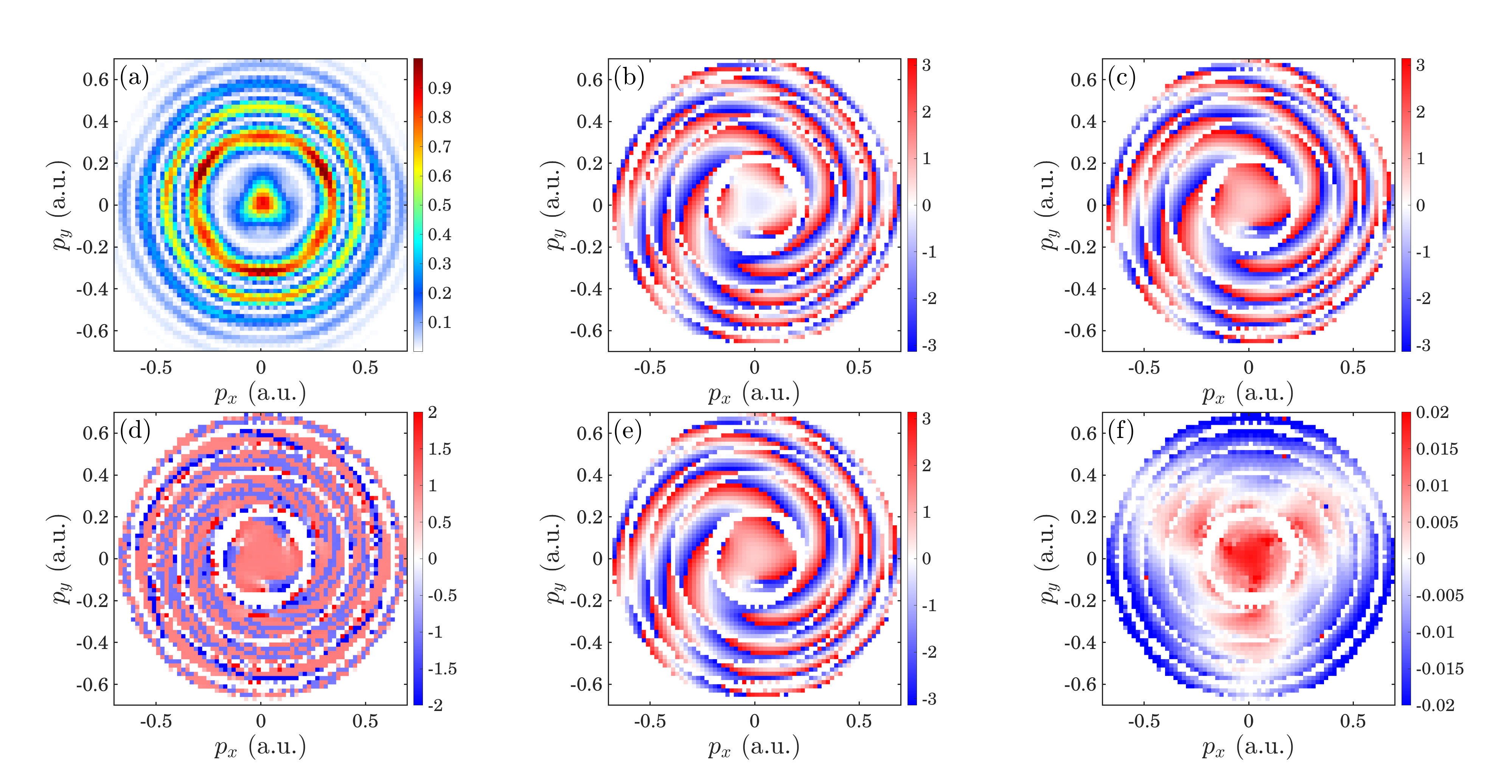}
	\caption{Intensity (a) and phase (b) of the photoelectron wavepacket ejected by a counter-rotating bicircular field calculated by SFA, in the $p_z=0.12$ a.u. plane. (c) Phase of the wavepacket in the $p_z=0.18$ a.u. plane. (d) Phase difference between these two planes. (e) Phase of the wavepacket (in rad) obtained by modulating the SFA signal by a chirosensitive momentum transfer due to the electron angular momentum at the exit of the tunnel [$\mathcal{M}=L_z(t_i')$], in the $p_z=0.18$ a.u. plane. (f). Phase difference between the modulated and unmodulated wavepacket in the $p_z=0.18$ a.u. plane. A mask is applied to the phase maps to hide the signal in areas where the photoelectron yield is below $5\%$ of its maximum.}
\label{FigVortex}
\end{center}
\end{figure*}

The difference between Fig. \ref{FigSimulCorot} and Fig. \ref{FigSimulCounter} confirms the importance of interference effects in strong field ionization by counter-rotating bicircular fields, and raises an interesting question: how can interference effects induce a rotation of the photoelectron momentum distribution, i.e. an attoclock shift? As we mentioned before, a phase shift between direct and indirect electrons emerging from the tunnel is expected to change the position of the interference phase jump, but cannot rotate the distribution away from  $\varphi_0=90^\circ$ or $150^\circ$ within the SFA. This means that an additional process is at play in chiral tunneling. To understand the origin of the chiral interferometric attoclock shift, we examine in Fig. \ref{FigVortex} the phase of the photoelectron wavepacket produced by the counter-rotating bicircular field. Fig. \ref{FigVortex}(b) shows a cut of the wavepacket phase in the $p_z=0.12$ a.u. plane. The phase presents a clear vortex structure, reflecting the temporal evolution of the action accumulated by the electrons in the continuum. For a given ATI peak, i.e. a given radius $p_x^2+p_y^2$, the angular phase evolution is characteristic of a spiral wavefront, carrying orbital angular momentum \cite{eickhoff2020,maxwell2020a}. Comparing the phase of the wavepacket obtained in two planes, $p_z=0.12$ (Fig. \ref{FigVortex}(b)) and  $p_z=0.18$ a.u. (Fig. \ref{FigVortex}(f)), reveals a high sensitivity of the vortex structure to the electron momentum, which can be visualized by plotting the phase difference between the two $p_z$ planes (Fig. \ref{FigVortex}(d)). In a pure SFA calculation, this phase difference has no effect, and the vorticity of the electron wavepacket is invisible in the electron momentum distribution, which is the modulus square of the wavepacket. However, if the chirality of the potential induces a $\delta p_z$ shift of the electrons emerging from the tunnel, then electrons corresponding to different $p_z$ in the pure SFA case end up at the same $p_z$. In that case, their different phases can lead to a rotation of the interference pattern. To quantify this effect, we plot in Fig. \ref{FigVortex}(e) the phase difference between the electron wavepacket calculated assuming $\mathcal{M}=L_z(t_i')$ and a pure SFA wavepacket. The phase difference is about 20 mrad, and shows a twisted angular structure. Thus, even if the chirosensitive modulation  $\mathcal{M}=L_z(t_i')$ is the same for electrons born on the leading and trailing edge of the field, and cannot produce an attoclock shift of each electron wavepacket, the phase twist resulting from the electron angular momentum in the continuum induces a rotation of the interference pattern, which appears as a chiral attoclock shift. Similarly, in the  $\mathcal{M}=C_\Re$ hypothesis, the opposite modulations of direct and indirect electrons does not result in a FBA in the yield of each wavepacket, but produces one in the interference pattern because of the phase vortex. 

\section{Conclusion and perspectives}
\label{section5}
We have shown that interference effects between electrons ejected on the leading and trailing edges of the field oscillations play a crucial role in photoionization by counter-rotating bicircular fields. In the strong-field ionization of chiral molecules, this interference has enabled us to increase the sensitivity of the attoclock scheme and to identify characteristic chirosenstive jumps along the electron ejection direction. The TDSE calculations in chiral molecules with a screened long-range potential have shown that this feature can be produced by short-range effects, i.e. be a consequence of non-adiabatic tunnel-ionization. We thus used SFA calculations to investigate the possible sources of chirosensitivity of the tunneling process. This study revealed the importance of non-adiabatic effects in the ionization by bicircular fields, induced by the field rotation as well as by its amplitude modulations. While some parameters of the complex sub-barrier electron motion, such as the electron exit angular momentum, are identical for direct and indirect electrons, others, such as the electron trajectory curvature, were found to switch sign. We tested the impact of these two categories of parameters on the chirosensitive response, and found that interference effects are determinant in shaping the attoclock angle and forward/backward asymmetry, because of the vorticity of the electron wavefront imposed by the rotating laser field. 

The interferometric chiral attoclock scheme thus enables to reveal the influence of the electron phase vortex in the strong field ionization. The twist, vorticity and angular orbital momentum of photoelectrons has been the subject of several recent studies.  Ngoko Djiokap \textit{et al.} have observed interference vortices in calculations of the photoionization by two delayed counter-rotating circularly polarized attosecond pulses \cite{ngokodjiokap2015}. Such vortices were measured by  Pengel \textit{et al.} in the multiphoton regime, using two delayed counter-rotating  laser fields \cite{pengel2017}. This configuration has been extensively studied theoretically in the strong field regime by Maxwell \textit{et al.} \cite{maxwell2020a}, using the SFA as well as more sophisticated models. In chiral molecules, Planas \textit{et al.} have calculated that strong field ionization by linearly polarized light produced enantio-sensitive electron vortices \cite{planas2022a}. These vortices are predicted to produce modifications of the electron-ion differential cross section \cite{tolstikhin2019}, opening new perspectives for the recently demonstrated chiral laser-induced electron diffraction \cite{rajak2024}.

\begin{figure*}
\begin{center}
\includegraphics[width=0.7\textwidth]{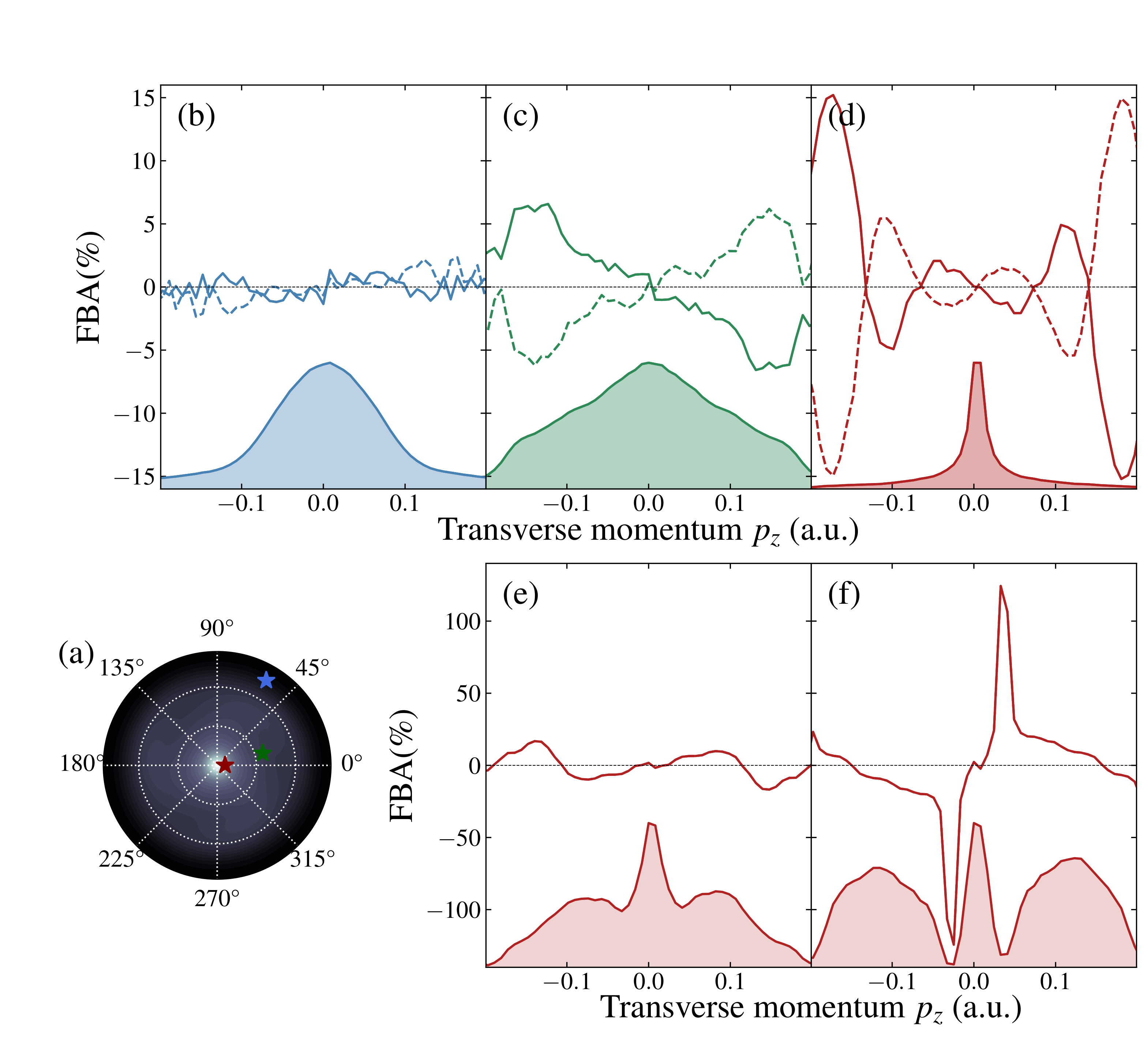}
	\caption{Experimental observation of chirosensitive transverse interference in (+)-camphor and (+)-fenchone, ionized with counter-rotating bicircular fields with right circular polarization of the fundamental. (a) Projection of the photoelectron angular distribution in the laser polarization plane, in (+)-camphor at $I=8\times 10^{12}$~W$\cdot$cm$^{-2}$ with $r=I_{2\omega}/I_{\omega}=1$. (b-d) Forward/Backward asymmetries in (+)-camphor (continuous lines) and (-)-camphor (dashed lines) and transverse momentum distributions (shaded areas) at $(p_x,p_y)=(0.02,0)$, $(0.11,0.03)$, and $(0.12,0.21)$ a.u., which correspond to the three spots marked in panel (a). (e) Same as (d), but at $I=1.4\times 10^{13}$~W$\cdot$cm$^{-2}$ with $r=0.4$. (f) Same as (e), but in (+)-fenchone. }
\label{FigInterf}
\end{center}
\end{figure*}

Another interesting aspect of the interferometric nature of photoionization in counter-rotating bicircular fields is its enhanced sensitivity to chirality. This sensitivity can be further increased by adjusting the intensity ratio $r$ between second harmonic and fundamental components. In the results presented above, this ratio was rather low (0.1), such that the dynamics of the two families of electron trajectories in the continuum was very similar. When this ratio is increased, the indirect electrons revisit the ionic core region after being released, ending up with an enhanced sensitivity to the influence of the ionic potential. This effect is further increased by the initial momentum of the electron at the exit of the tunnel \cite{kaushal2018,geyer2023}. In that situation, we can expect an increased structural sensitivity of the interference between direct and indirect electrons. Eckart \textit{et al.} have established that this interference could be studied by monitoring the electron momentum distribution along the light propagation direction. Figure \ref{FigInterf}(a) shows measurements of the projection of the low-energy part of the distribution onto the polarization plane, for photoionization of camphor molecules by a counter-rotating bicircular field at $I=8\times 10^{12}$~W$\cdot$cm$^{-2}$ and $r=1$. When the  longitudinal momentum $(p_x,p_y)$ is large, the transverse $p_z$-distribution displayed in Fig. \ref{FigInterf}(b) has a symmetric Gaussian shape. The FBA is close to zero. The electrons ejected in this range are thus insensitive to the chirality of the molecular potential. In the region of intermediate longitudinal momentum, exemplified by $(p_x,p_y)=(0.11,0.03)$ a.u. in Fig. \ref{FigInterf}(c), modulations of the transverse distribution appear, reflecting the interference between direct and indirect electrons. These modulations are forward/backward asymmetric, and thus sensitive to chirality. The FBA reaches $7\%$, and reverses perfectly when switching the enantiomer to (-)-camphor. At very low longitudinal momentum $(p_x,p_y)\approx(0,0)$, the transverse distribution is sharply peaked about $p_z=0$, indicating a strong Coulomb focusing (Fig. \ref{FigInterf}(d)). Coulomb focusing was theoretically shown to increase the sensitivity of photoionization to molecular chirality \cite{rozen2021}. This is confirmed by these measurements, which show that the momentum distribution is highly asymmetric along $p_z$, with a FBA oscillating as $p_z$ varies and reaching 15$\%$. These oscillations are the signature of the interference between direct and indirect electrons. Increasing the laser intensity to $I=1.4\times 10^{13}$~W$\cdot$cm$^{-2}$ with $r=0.4$ leads to an increase of the contribution of the rescattering electrons, which clearly appear on the transverse momentum distribution (Fig. \ref{FigInterf}(e)), with a FBA reaching 17$\%$. Last, to illustrate the structural sensitivity of the scheme, we repeated the measurements in fenchone. The asymmetric interference pattern is even more contrasted. The interference between direct and delayed electrons is almost fully destructive in the backward direction, but not in the forward one. This leads to a very high value of FBA, reaching 120$\%$ -- the maximum possible FBA value being 200$\%$ with the definition employed. 

Last, it would be interesting to investigate the importance of the sub-barrier dynamical parameters, such as the trajectory curvature, on spin-polarization effects \cite{barth2013,hartung2016}, in the line of the pioneering work carried out by Kaushal and Smirnova \cite{kaushal2018,kaushal2018a,kaushal2018b} and Ayuso \textit{et al.}\cite{ayuso2017}. Measuring spin-resolved quantum interference patterns arising in counter-rotating fields could provide an interesting approach to probe dynamical spin effects in strong-field ionization. 

We thank Franck Blais, Rodrigue Bouillaud and Laurent Merzeau for technical assistance. We acknowledge David Ayuso, Val\'erie Blanchet and Olga Smirnova for fruitful discussions. This project has received funding from the Agence Nationale de la
Recherche (ANR) -- Shotime (ANR-21-CE30-038-01), and from the European Research Council (ERC) under the European Union's Horizon 2020 research and innovation programme no. 682978 - EXCITERS. Funded by the European Union (ERC Starting Grant UTOPIQ ERC-2022-STG No.101076639). Views and opinions expressed are however those of the authors only and do not necessarily reflect those of the European Union. Neither the European Union nor the granting authority can be held responsible for them.

\bibliographystyle{apsrev4-1}
%\bibliography{biblio.bib}

%

\end{document}